\documentclass{emulateapj}
\usepackage{graphicx}
\shorttitle{Low Frequency QPOs in Black-Hole Candidates}
\shortauthors{P.Casella et al.}
\begin{document}
\def\new#1{{#1}}
\title{The ABC of low-frequency quasi-periodic oscillations in Black-Hole Candidates: analogies with Z-sources}

\author{P. Casella\altaffilmark{1,2}, T. Belloni\altaffilmark{1} 
and L. Stella\altaffilmark{2}}
\altaffiltext{1}{INAF - Osservatorio Astronomico di Brera,
              via E. Bianchi 46, I--23807 Merate (LC), Italy; casella@merate.mi.astro.it}
\altaffiltext{2}{INAF - Osservatorio Astronomico di Roma,
       Via di Frascati, 33, I--00040 Monteporzio Catone (Roma), Italy}
\thispagestyle{empty}
\begin{abstract}
Three main types of low-frequency quasi-periodic oscillations (LFQPOs) have
been observed in Black Hole Candidates. We re-analyzed RXTE data of the bright
systems XTE J1859+226, XTE J1550-564 and GX 339-4, which show all three of
them. We review the main properties of these LFQPOs and show that they follow
a well-defined correlation in a fractional rms vs. softness diagram. We show
that the frequency behavior through this correlation presents clear analogies
with that of Horizontal-, Normal- and Flaring-Branch Oscillations in Z
sources, with the inverse of the fractional rms being the equivalent of the
curvilinear coordinate $S_z$ through the Z track.
\end{abstract}


\section{Introduction}

Low-frequency Quasi-Periodic Oscillations (LFQPOs) with centroid frequencies
from mHz to tens of Hz have been observed in the X-ray flux of many
neutron-star and black-hole X-ray binaries
\citep[see][and references therein]{vdK05}. 
In neutron-star Low Mass X-ray Binaries (LMXRB) the timing behavior has long 
been known to correlate with spectral variations. In particular, the
properties of the LFQPOs vary in systematic fashion along the pattern that
these sources describe in the X-ray color-color diagram (CD).
In high-luminosity neutron-star systems (the so-called Z sources) 
\citep{HK89,vdK05}, three types of LFQPOs have been associated with the
position along the Z-pattern that these sources describe in 
a CD: the horizontal branch oscillations (HBOs), normal branch 
oscillations (NBOs) and flaring branch oscillations (FBOs) \citep[for details
on these LFQPOs, and how they are tracked through the Z-pattern by using the
curvilinear coordinate $S_z$ see][]{vdK95}. 

In the case of Black Hole Candidates (BHCs) the general picture is at present
less clear. Three main types of LFQPOs,
dubbed Type-A, -B and -C respectively, originally identified in the light
curve of XTE J1550-564 \citep[see][]{Wijnandsetal99a,Remillardetal02}, have
been seen in several sources (see Section 2). On the other hand, three main
bright states (in addition to the quiescent state) have been identified in
these sources, based on their spectral and timing properties \citep[for a
review see][]{Tanaka&Lewin95,vdK95,McC&R05,vdK05,H&B05}. 
It was only very recently that systematic variations in the energy spectra and
intensity of transient BHCs could be identified in terms of the pattern
described in an X-ray Hardness-Intensity diagram (HID)
\citep[see][]{Homanetal01,Belloni03,H&B05,Bellonietal05}. 
Original states are found to correspond to different branches/areas of a
square-like HID pattern. In this scheme the BHC LFQPO phenomenon appears to be
confined to within a comparatively small range of spectral properties of these
sources, requiring a deeper investigation over a restricted range in
parameters' space. Within this small range, attempts to correlate the QPO
properties  with other properties such as the source position in the CD or the
HID across different BHCs have so far given inconclusive results. This is
clearly at variance with neutron-star LMXRBs. 

\new{Similarities between power spectra in BHCs and in Z sources have been
stressed by several authors \citep[see e.g.][]{Miyamotoetal93}. \citet{vdK94}
discussed the parallelism between NS and BH systems and underlined the
importance of quantitatively studying the similarities between them.
\citet{W&vdK99a} and \citet{PBK99} found global correlations between
characteristic frequencies in both NS and BH systems, involving type-C
LFQPOs and HBOs for BHCs and Z sources respectively, plus the low-frequency
LFQPOs in atoll sources. These correlations suggested that basic frequencies
of these systems likely have the same origin as envisaged in some QPO models,
and yielded to propose an association between type-C LFQPOs and
HBOs.}

In this letter, we show that the QPO type and centroid frequency 
in  BHCs vary systematically as a function of the inverse of 
the source rms fractional variation. This behavior is reproduced fairly 
accurately over different BHCs and presents clear similarities 
with the LFQPOs of neutron star low mass X-ray binaries. 
Based on this analogy we suggest that C, B and A type LFQPOs in
BHCs correspond to HBOs, NBOs and FBOs of high luminosity
neutron-star systems of the Z-class respectively. 


\section{QPO classification} \label{sect3Types}

Several distinct types of LFQPOs showing different properties have been 
discovered in BHCs. 
An exhaustive classification has not yet been obtained, given the
complexity and variety of the observed behaviors. However, three main LFQPO
types (named Type-A, -B and -C) stand out in the present scenario. In Table
\ref{qpotable} we summarize their main properties.

\subsection{Type C LFQPOs} \label{sectTypeC}

\begin{deluxetable}{lccc}[b]
  \tabletypesize{\scriptsize}
  \tablecaption{Summary of type-A, -B and -C LFQPOs properties\label{qpotable}}
  \tablewidth{0pt}
  \tablehead{
    \colhead{Property} & \colhead{Type C} & \colhead{Type B} & \colhead{Type A}
  }
  \startdata
            Frequency (Hz) & $\sim$0.1-15 & $\sim$5-6 & $\sim$8 \\
            Q ($\nu$/FWHM) & $\sim$7-12 & $\ga$6 & $\la$3 \\
            Amplitude (\%rms) & 3-16 & $\sim$2-4 & $\la$3 \\
            Noise & strong flat$-$top & weak red & weak red \\
            Phase lag @$\nu_{QPO}$ & soft/hard & hard & soft \\
            Phase lag @2$\nu_{QPO}$ & hard & soft & ... \\
            Phase lag @$\nu_{QPO}$/2 & soft & soft & ... \\
  \enddata
   \end{deluxetable}

Type-C LFQPOs are characterized in the power spectrum by a strong \new{(up to
 $\sim$16\% rms)}, narrow \new{($\nu/\Delta\nu \sim 7-12$)} and variable peak \new{(its centroid frequency and intensity varying by several percent in
 a few days, see e.g. C04)} at frequencies $\sim$0.1-15 Hz, superposed on a
Flat-Top Noise (FTN) that steepens above a frequency comparable to the QPO
frequency \citep[see][]{W&vdK99a,BPK02}. The total (QPO and FTN) 
fractional rms variability can be as high as $\sim$ 30\%. 
A subharmonic and a second harmonic peak are often seen. Phase lags
(i.e. the phase delay between two light curves at different energies) depend
strongly on the frequency of the QPO, with a trend towards soft lags
(i.e. soft X-ray variations lag those at hard X-ray energies) for
increasing QPO frequency \citep[see][]{Reigetal00}, but they are usually
consistently soft at the subharmonic and hard at the second
harmonic \citep[see e.g.][C04 hereafter]{Remillardetal02,C04}. The
QPO rms increases with energy, flattening above $\sim$ 10 keV (see e.g. C04). 
In some cases a decrease above 20 keV is observed \citep{Tomsick&Kaaret01}
which might be associated with a higher radio flux \citep{Rodriguezetal04}.

Low-frequency QPOs that can be identified as type-C were observed
in a number of sources, e.g. GS 1124-684
\citep[]{Miyamotoetal93,Takizawaetal97}, XTE J1550-564
\citep{Remillardetal02}, XTE J1859+226 (C04), GX 339-4 \citep{Miyamotoetal91}
and GRO J1655-40 \citep{Mendezetal98}.

The presence of a strong FTN component and the correlation of the QPO
frequency with the source intensity and the FTN break frequency
\citep{W&vdK99a} strongly suggest the association of this low frequency QPO
type with the HBO observed in Z sources \citep[]{Miyamotoetal93}.

\subsection{Type B LFQPOs} \label{sectTypeB}

Type-B LFQPOs are characterized by a \new{relatively strong ($\sim$4 \% rms)
and narrow ($\nu/\Delta\nu \ge 6$)} peak which is found in a narrow range of
centroid frequencies \new{around 6 Hz}. Unlike type C, there is no
evidence for FTN, although a weak red-noise (few \% rms) is detected at very
low frequencies ($\la$ 0.1 Hz). A weak
second harmonic is often present, sometimes together with a subharmonic
peak. In a few cases, the subharmonic peak is higher and narrower (resulting
in a ``Cathedral-like'' shape, see
C04). Phase lags are hard at the fundamental and soft
at the second harmonic, i.e. the opposite of the behavior observed for type-C
LFQPOs. 
However, phase lags are soft at the subharmonic as in type-C LFQPOs. 
The energy dependence of the QPO rms is similar to
that of type-C LFQPOs, but the rms values are systematically lower
(factor $\sim 2$).

The presence of type-B LFQPOs has been reported in different sources: see
e.g. GS 1124-684 \citep[]{Miyamotoetal93,Takizawaetal97}, XTE J1550-564
\citep{Wijnandsetal99a}, GX 339-4 \citep{Nespolietal03,Bellonietal05}, XTE
J1859+226 (C04) and GRS 1739-278 \citep{Wijnandsetal01}. 

Rapid transitions in which type-B LFQPOs appear/disappear are often observed
in some of these sources. The
transitions are unresolved at present, as they take place on a time scale
shorter than a few tens of seconds. In GX 339-4, the QPO appearance is related
to spectral hardening of the source flux (see Nespoli et al 2004; Belloni et
al. 2005), while in XTE J1859+226 transitions towards type-C and type-A LFQPOs
appear to be related to a spectral softening and a spectral hardening,
respectively (C04). \new{A rapidly transient $\sim$6 Hz QPO was observed
also in the atoll source 4U 1820-30, and was compared with both NBOs observed
in Z sources \citep{Wijnandsetal99b} and type-B LFQPOs observed in BHCs
\citep{1820} (other NBO-like LFQPOs have been observed also in the atoll
sources XTE J1806-2646 \citep{W&vdK99b,Revnivtsevetal99} and Aql X-1
\citep{Reigetal04}).}

The centroid frequency of type-B LFQPOs (which shows often marked variability
on a time scale of $\sim$ 10~s, see Nespoli et al. 2003) is close to the
frequency range ($\sim 5-8$~Hz) over which NBOs are observed in neutron-star
systems of the Z-class (and perhaps also a few lower luminosity systems, see
van der Klis 1995). This frequency coincidence, together with the
fact that both type-B LFQPOs and NBOs show a low amplitude noise, seem to
suggest an association of these two LFQPO types. However, it must
be noticed that some important differences still remain between the two LFQPO,
\new{as} the higher coherence of type-B LFQPOs \new{and the lack of harmonic
content in NBOs. Moreover, NBOs are seen simultaneously with HBOs
in Z sources \citep[see][and references therein]{vdK95}, demonstrating they
are different phenomena. This does not happen in BHCs, where type-B LFQPOs
are seen to switch from/to type-C LFQPOs (see previous paragraph) without any
contemporaneity between the two.
}
Furthermore, the
centroid frequency of type-B LFQPOs in XTE J1859+226 shows a weak positive
correlation with the count rate \citep{cospar04}. It is thus in principle
possible that the $\sim$1~Hz QPO observed in GX 339-4 at low count rates
\citep{Bellonietal05} could be identified as a type-B. This would extend the
frequency range where type-B LFQPOs are observed in BHCs.

   \begin{figure}
     \includegraphics[width=8.0cm]{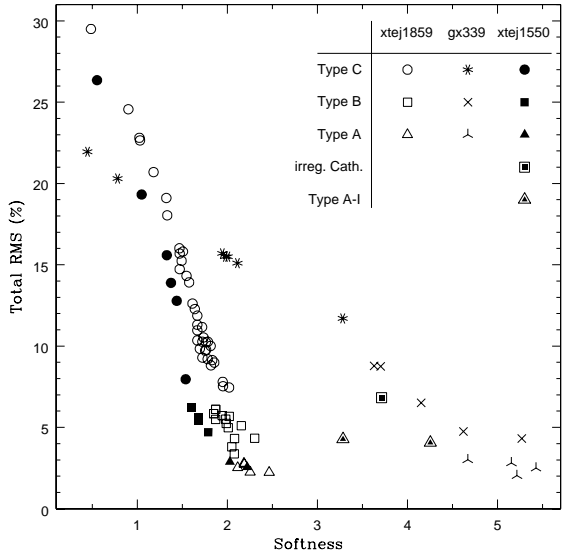}
     \caption{Total 0.06-64 Hz rms (in the 2-15 keV band) versus softness
     (defined as the ratio between $\sim$2-3.4 keV and $\sim$9-20 keV) for
     those observations in which a low frequency QPO was observed. Different
     symbols are explained in the inset table (see text for the irregular
     types).}
     \label{rette}
   \end{figure}


\subsection{Type A LFQPOs} \label{sectTypeA}

Type-A LFQPOs are characterized by a weak \new{(few \% rms)} and broad
\new{($\nu/\Delta\nu \le 3$)} peak around 8 Hz. A
very low amplitude red noise is observed, whereas neither a subharmonic nor
a second harmonic were present \new{(possibly because of the width of the
fundamental peak)}. This LFQPO was observed in different
sources:  GS 1124-684 \citep[]{Miyamotoetal93,Takizawaetal97}, GX 339-4
\citep{Nespolietal03,Bellonietal05} and XTE J1859+226 (C04). C04 showed that
phase lags at the frequency of the QPO are soft, while they were not
measurable in the rest of the frequency range because of poor
statistics. In some sources a deeper analysis is
necessary in order to confirm these identifications.

At first, these LFQPOs were dubbed ``type A-II'' by \citet{Homanetal01}. LFQPOs
 dubbed as ``type A-I'' \citep{Wijnandsetal99a} were strong, broad,
and accompanied by a very low-amplitude red noise. Moreover, a shoulder on the
 right hand side of this QPO was clearly visible and interpreted as a very
broadened second harmonic peak. In Section \ref{sectCorrel} we discuss that
``type A-I'' LFQPOs should be classified as type B.

The higher frequency and lower coherence of type-A LFQPOs, relative to
type-B LFQPOs, is suggestive of an analogy with the FBOs observed in
Z sources. Moreover, as in Z-sources the flaring branch is effectively
``flaring'' only in some sources \citep{vdK05}, also in BHCs the soft
observations where type-A LFQPOs appear correspond only in some sources to the
highest flux values \citep[see e.g. C04 and][]{Nespolietal03}. However, there
are also some
differences. First, the rms amplitude of FBOs is usually
stronger than that of type-A LFQPOs. Second, rapid (tens of seconds) and
unresolved transitions appear to take place between type
A and B LFQPOs \citep[see C04 and][]{Nespolietal03}. On the contrary, in the
case of Z sources fairly continuous evolution is clearly observed in the
transition between NBO and FBO \citep{Dieters&vdK00}.

\section{Correlations} \label{sectCorrel}

The three LFQPO types described in Section \ref{sect3Types} are in most cases
observed in the top branch of the HID (see Belloni et al. 2005). 
In the three sources which have unambiguously shown all three types of
LFQPOs (XTE J1550-564, XTE J1859+226 and GX 339-4), type-Bs appear at
lower hardness than type-Cs and higher hardness than type-As, thus identifying
a well definite sequence C$\rightarrow$B$\rightarrow$A.

   \begin{figure}
     \includegraphics[width=8.0cm]{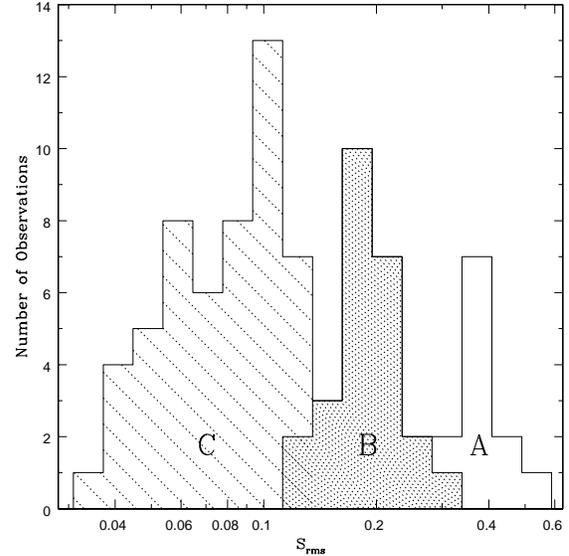}
     \caption{Histogram of the S$_{rms}$ values. Different grayscales indicate
     different LFQPO types.}
     \label{histogram}
   \end{figure}


However, given the lack of easily identifiable geometrical points of reference,
it has not yet been possible to track unambiguously the evolution of LFQPOs
properties along the HID as it has been done with the CDs in neutron
stars. Perhaps this is
due to the fact that, given the softness of the BH spectrum, the
most dramatic spectral changes take place shortwards of $\sim$1-2 keV, where
present instrumentation does not permit a detailed characterization of
the small region of the HID where the different types of LFQPOs are observed.

In order to find a parameter unambiguously tracking the
type-C$\rightarrow$B$\rightarrow$A sequence, we plot the
total integrated rms (0.06-64 Hz, in the 2-15 keV energy range) vs. the
softness (defined as the ratio between the PCA counts in the energy bands
$\sim$2-3.4 keV and $\sim$9-20 keV) of the observations where one of the three
LFQPO types was observed in each of the three above-mentioned sources (see
Fig. \ref{rette}). For XTE J1859+226 and GX 339-4, we used the complete
samples, using
classifications reported respectively by C04 and Belloni et al. (2005). For
XTE J1550-564, we chose all type-A and -B LFQPOs for which the
identification was incontrovertible, and several type-C LFQPOs covering the
observed frequency range (identifications from \citet{Wijnandsetal99a},
\citet{Homanetal01} and \citet{Remillardetal02}). In order to obtain values of
rms and hardness consistent among the three sources, we re-analyzed all the
data (for technical details on the analysis we refer to C04).

The behavior of the three sources in Fig. \ref{rette} is similar: the
softness and the rms show a monotonic, roughly linear anti-correlation between
each other, through which the C-B-A sequence is well defined. 
Only three points deviate from this scheme, all of them from XTE J1550-564
(framed square and triangles, see the inset
table). In these three cases, the identification of the LFQPO is uncertain: one
is a ``B-Cathedral'' type with an unusual band-limited noise,
the other two were classified as type A-I (see Section \ref{sectTypeA} and in
the following). It is worth mentioning that in the case of XTE J1859+226,
the points where no LFQPOs could be clearly identified (see
C04) follow a different correlation in the rms vs. softness diagram, lying on a
flatter linear-like track (not shown in Fig. \ref{rette}) at higher values of
softness, and that the same happens in the case of GX 339-4
\citep{Bellonietal05}. 

However, to study in detail the complete evolution of the
outburst in the rms-softness diagram is beyond the scope of this work, in
which we concentrate on the study of the unambiguously classified LFQPOs.

While the range in softness differs from one source to another (see in
particular the points for GX 339-4) it is apparent from Fig. \ref{rette} that
all three sources span a similar range in rms fractional variability. This
suggest that the rms fractional variability can be used as a parameter
tracking the three LFQPO types. In the histogram of Fig. \ref{histogram} we
use a related parameter, the inverse of the rms (hereafter $S_{rms}$), in
order to maintain the C-B-A sequence observed in the HID.
It is evident that the points of the three sources cluster
around three values of $S_{rms}$, with the three groups corresponding to the
three LFQPO types reviewed in Section \ref{sect3Types}.

   \begin{figure}
     \includegraphics[width=8.0cm]{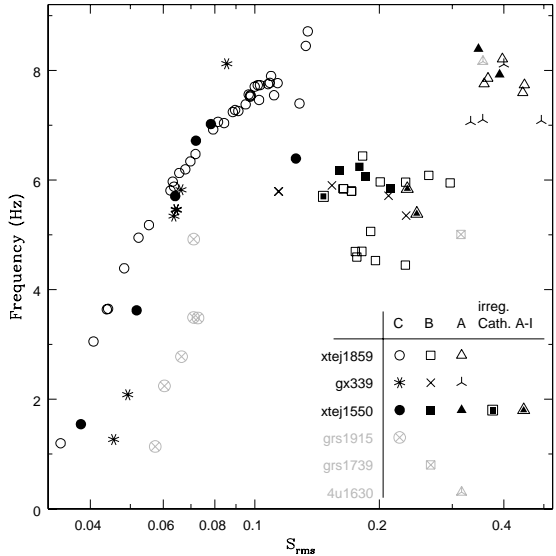}
     \caption{QPO frequency as a function of $S_{rms}$. Different symbols are
     explained in the inset table (see text for the irregular types).}
     \label{musl}
   \end{figure}


   \begin{figure}
     \includegraphics[width=8.0cm]{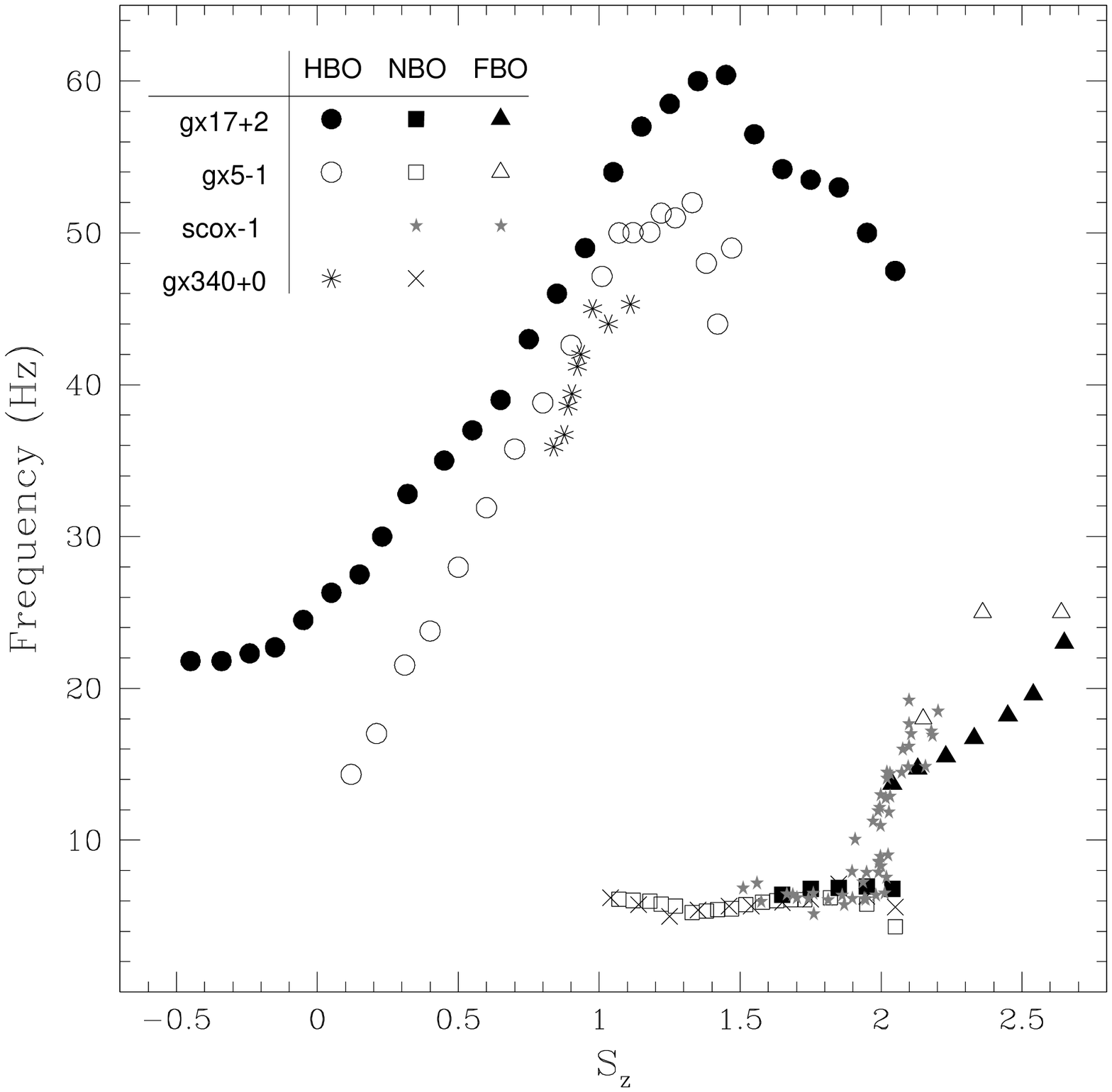}
     \caption{QPO frequency through the Z tracks of GX 17+2
     (data from Homan et al. 2002), GX 5-1 \citep{Jonkeretal02}, Sco X-1
     \citep{Dieters&vdK00} and GX 340+0 \citep{Kuulkers&vdK96,Jonkeretal00}.}
     \label{musz}
   \end{figure}


In Fig. \ref{musl} we plot the frequency of the LFQPOs versus
$S_{rms}$. The separation among three LFQPO types is well defined, and
data from the three sources overlap fairly accurately, creating
a unique characteristic shape. The frequency of type-C LFQPOs increases steeply
up to $\sim$ 8 Hz at $S_{rms} \sim$ 0.15. After this point, type-B
LFQPOs appear with frequencies (mainly) close to $\sim$ 6 Hz until $S_{rms}
\sim$
0.3. Finally, for higher values of $S_{rms}$, only type-A LFQPOs are present,
with frequencies around 8 Hz. It is worth noting that the three LFQPOs
which deviated from the correlation in Fig. \ref{rette} lie now in the
region of the plot where type-B LFQPOs are concentrated. If in the case of the
``Cathedral'' LFQPO, this is not surprising, in the other two cases this seems
to be either an exception to the separation among the three LFQPO types or
a problem with the ``type A'' definition of the QPO themselves. On the other
hand, the clustering of type A-I LFQPOs around $\sim$6 Hz \citep{Homanetal01}
clearly makes them standing out among type-A LFQPOs. The strong difference in
amplitude between type A-I and A-II leaves room for a separation of
the two classes and for an association of the type A-I with the type
B. In Fig. 3 we also plot data from a few other
sources (see inset table) which show only one of the three LFQPO types,
but clearly confirm the general scheme.

The behavior in Fig. \ref{musl} is strongly reminiscent of the scheme observed
in the Z sources, with the sequence $HBO \rightarrow NBO \rightarrow FBO$
along the Z track. In Fig. \ref{musz} we plot the QPO frequency versus the
rank number $S_z$ of four Z sources. The similarity with Fig. \ref{musl} is
evident and further supports the association between Z sources' and
BHCs' LFQPOs (see Sect. 2).


\section{Conclusions}

We reviewed the evidence that BHCs present three main different types
of Low-Frequency QPOs. Each of these three types has well defined properties
and shows strong similarities with one of the three types of LFQPOs observed
in Z sources. 
The three types appear in a well defined sequence along the HID, as the three
types of LFQPO observed in Z sources do along the CD.

Furthermore, we found that their frequency follows a characteristic trend 
\new{as a function of the total integrated fractional variability}. 
This trend is clearly reminiscent of that
observed along the Z track of Z sources. On the basis of these
similarities we propose to associate the C, B and A types respectively to the
HBO, NBO and FBO observed in the Z sources\new{, thus strengthening and
extending the previously proposed associations}. In
Table \ref{QPOproperties} we summarize the main similarities in the properties
of these LFQPOs.

If these associations are correct, than the Type-Cs in BHCs and the HBOs in
Z sources might well be caused by a similar physical mechanism. The fact
that the frequencies of the type-Cs are roughly a factor of $\sim$6-7 lower
than the frequencies of the HBOs suggests that these frequencies scale
approximately as the inverse of the mass of the compact object, as expected for
dynamical frequencies. On the other hand, both NBOs and type-B LFQPOs have
frequencies close to 6 Hz, thus it is natural to suppose that the physical
mechanism that determines their frequency is independent of (or only weakly
dependent on) the mass of the compact object.

The presence of these two mechanisms in both types of
compact objects would rule out all models that involve any interaction with
the surface or the magnetosphere of the neutron star. 

\begin{deluxetable}{lcc}[h]
\tabletypesize{\scriptsize}
\tablecaption{Summary of the main properties of type-C and type-B LFQPOs
compared with HBOs and NBOs.\label{QPOproperties}}
\tablewidth{0pt}
\tablehead{
\colhead{Property} & \colhead{HBO / Type C} & \colhead{NBO / Type B}
}
\startdata
            Frequency (Hz) & intensity dependent & 5$\sim$6 Hz \\
            Amplitude (rms) & anti$-$correlated & 2-4 \% \\
             & with frequency &  \\
            Noise & strong flat$-$top & weak red \\
\enddata
\end{deluxetable}


\acknowledgments

This work was partially supported by MIUR under CO-FIN grants 2002027145 and
2003027534.


\end{document}